\documentclass[letterpaper,american,9pt,twocolumn,twoside]{revtex4-1}
\usepackage[T1]{fontenc}
\usepackage[utf8]{inputenc}
\usepackage{amsmath}
\usepackage{amssymb}
\usepackage{graphicx}
\usepackage{esint}

\makeatletter

\pdfpageheight\paperheight
\pdfpagewidth\paperwidth

\@ifundefined{textcolor}{}
{%
 \definecolor{BLACK}{gray}{0}
 \definecolor{WHITE}{gray}{1}
 \definecolor{RED}{rgb}{1,0,0}
 \definecolor{GREEN}{rgb}{0,1,0}
 \definecolor{BLUE}{rgb}{0,0,1}
 \definecolor{CYAN}{cmyk}{1,0,0,0}
 \definecolor{MAGENTA}{cmyk}{0,1,0,0}
 \definecolor{YELLOW}{cmyk}{0,0,1,0}
}

\usepackage{diagbox}
\usepackage{makecell}
\usepackage{upgreek}
\AtBeginDocument{
\addtolength{\abovedisplayskip}{-2ex}
\addtolength{\abovedisplayshortskip}{-2ex}
\addtolength{\belowdisplayskip}{-2ex}
\addtolength{\belowdisplayshortskip}{-2ex}
}

\makeatother

\usepackage{babel}
\begin{document}

\title{Spatial Coherence and the Orbital Angular Momentum of Light in Astronomy}

\author{D. Hetharia}

\affiliation{Huygens--Kamerlingh Onnes Laboratory, Leiden University, P.O. Box
9504, 2300 RA Leiden, The Netherlands}

\author{M. P. van Exter}

\affiliation{Huygens--Kamerlingh Onnes Laboratory, Leiden University, P.O. Box
9504, 2300 RA Leiden, The Netherlands}

\author{W. Löffler}

\email{loeffler@physics.leidenuniv.nl}

\affiliation{Huygens--Kamerlingh Onnes Laboratory, Leiden University, P.O. Box
9504, 2300 RA Leiden, The Netherlands}
\begin{abstract}
The orbital angular momentum (OAM) of light is potentially interesting
for astronomical study of rotating objects such as black holes, but
the effect of reduced spatial coherence of astronomical light sources
like stars is largely unknown. In a lab-scale experiment, we find
that the detected OAM spectrum depends strongly on the position of
the light-twisting object along the line of sight. We develop a simple
intuitive model to predict the influence of reduced spatial coherence
on the propagating OAM spectrum for, e.g., astronomical observations.
Further, we derive equations to predict the effect of line-of-sight
misalignment and the received intensity in higher-order OAM modes
for limited-size detectors such as telescopes. 
\end{abstract}


\maketitle

The total angular momentum of paraxial light fields contains a spin
(polarization) and an orbital part, where the latter is related to
the azimuthal component of light's spatial degree of freedom. For
rotationally invariant intensity distributions, there are ``pure''
orbital angular momentum (OAM) fields that are characterized simply
by a helical phase $e^{i\ell\phi}$, where $\phi$ is the azimuth
in the chosen coordinate system, and $\ell\hbar$ is the OAM of a
single photon in such a mode \cite{allen1992}. In general, we can
characterize light via its OAM spectrum $P_{\ell}$ \cite{molinaterriza2001}.
The OAM of light is proven to be useful in a broad range of classical
and quantum optical applications, and plays a key role in vortex coronagraphy
\cite{foo2005,mawet2005} in astronomy, but it is an open question
wether the OAM of light from deep space objects is useful for astronomical
observations on earth \cite{harwit2003,berkhout2008,berkhout2009}.
There is potential, as for instance, frame-dragging of space in the
vicinity of a rotating (Kerr) black hole can leave a significant trace
in the OAM spectrum of the light passing through the region \cite{carini1992,feng2001,tamburini2011,yang2014oam}.
This gives potentially direct access to the spin of black holes, which
is up to now only accessible via indirect methods such as relativistic
line broadening by the linear Doppler shift and gravitational effects
\cite{fabian2000,risaliti2013}. We note that there are also OAM phase
modifications of light passing through \cite{gotte2007,leach2008}
or being reflected from \cite{lavery2013rot} rotating objects due
to the rotational Doppler effect; but here, we study situations where
the \emph{total OAM} is modified and becomes non-zero. 

\begin{figure}
\includegraphics[width=1\columnwidth]{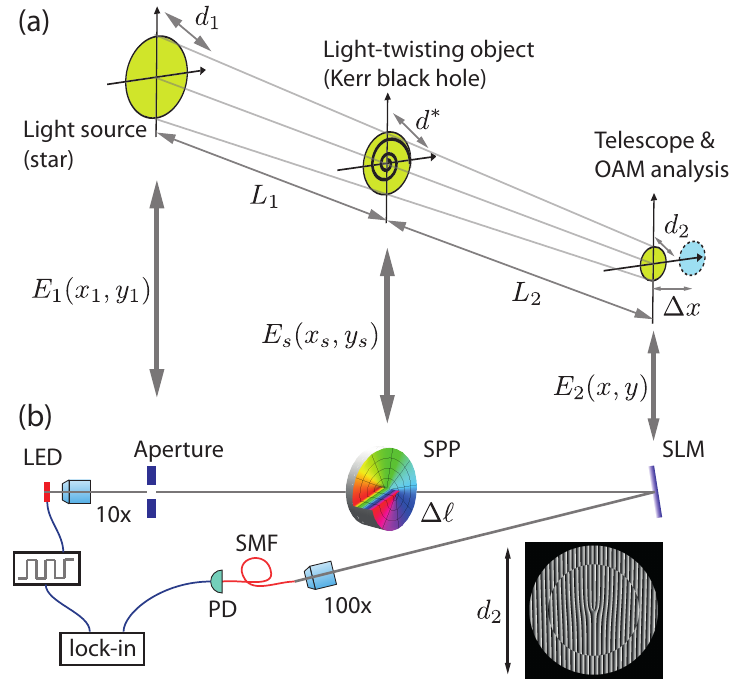}

\protect\caption{\label{fig:scheme}(a): The scheme: A spatially incoherent source
of diameter $d_{1}$ such as a star illuminates a region of space
that modifies the OAM of light, for instance a Kerr black hole. On
earth, a telescope or interferometer with an effective diameter (baseline)
of $d_{2}$ is used to measure the OAM spectrum. (b): Lab-scale experimental
setup. To simulate the star, we use a spatially incoherent light source
(LED) illuminating the first aperture; the spiral phase plate (SPP)
with charge $\Delta\ell$ mimics the light-twisting object that modifies
the OAM. A combination of phase-only spatial light modulation (SLM;
see inset for an exemplary hologram) and imaging onto the core of
a single-mode fiber (SMF) coupled to a photo-diode (PD) is used to
measure the OAM spectrum; the aperture $d_{2}$ on the SLM corresponds
to the telescope diameter.}
\end{figure}

Because the OAM of light is connected to the spatial degrees of freedom
of light, observation thereof requires light fields with sufficient
spatial (transverse) coherence \cite{mandel1995,born1999}. This is
why light emission from, e.g., the black hole accretion disk itself
cannot be used, instead, we consider the case of a star illuminating
the black hole from behind (Fig.~\ref{fig:scheme}a). But also stars
emit spatially incoherent light which acquires spatial coherence only
upon propagation; and even if we observe starlight on earth fully
coherent, it it might be incoherent at the light-twisting object.
The study of optical phase singularities in partially coherent fields
started with the static case of a twisted Gaussian Schell-Model beam
\cite{simon1993,serna2001}. Only relatively recently, the dynamic
case including propagation was investigated; this led to the discovery
of circular correlation singularities \cite{maleev2004,palacios2004,pires2010b,yang2013coh},
further, the precise structure of the vortices turned out to be quite
different to the coherent case \cite{swartzlander2007}. However,
the influence of reduced spatial coherence on the experimentally accessible
OAM spectrum for a light-twisting object in between is still largely
unknown; mostly light without total OAM was investigated so far \cite{pires2010b,malik2012}.
Because the light-twisting object modifies azimuthal correlations,
theoretical calculation of the propagation of the cross-spectral density
function are very time consuming \cite{swartzlander2007,pires2010b};
simple models are missing for, e.g., assessment of the situation in
astronomy \cite{harwit2003,tamburini2011}. We provide here firstly
such a model and confirm it by lab-scale experiments.

To set the stage we show in Fig.~\ref{fig:scheme} our scheme and
the lab-scale experiment emulating the source star, the light-twisting
object, and the detector. We consider here the case where those objects
lie approximately on a straight line, and spectrally coherent quasi-monochromatic
light. Further, we work within the paraxial approximation and assume
homogeneous polarization, therefore we discuss only scalar fields.
Astronomical light sources are nearly always spatially completely
incoherent, because of spatially uncorrelated light generation processes.
To synthesize the spatially incoherent light source (star) in the
lab, we image with a $10\times$ microscope objective a suitable spot
from a large-area LED chip ($\lambda=620\pm10$~nm) onto an aperture
of diameter $d_{1}$. During propagation to the light-twisting object
(at distance $L_{1}$), a certain degree of spatial coherence is build
up according to the van Cittert--Zernike theorem. The object then
imprints OAM, where we use in the experiment a spiral phase plate
(SPP) of charge $\Delta\ell$ \cite{khonina1992,beijersbergen1994}.
Finally, light propagates over distance $L_{2}$ to the observer,
a OAM-spectrum analyzer. This consists of a phase-only spatial light
modulator (SLM) whose surface is imaged with a $100\times$ microscope
objective onto the core of a single-mode fiber (SMF) connected to
a femtowatt photo detector. We modulate the LED at around $500\,\mathrm{Hz}$
and use lock-in detection. The SLM holograms are restricted to a circular
area to select our detection aperture $d_{2}$, which corresponds
to the telescope entrance aperture. For determination of the azimuthal-only
OAM spectrum, we need to integrate over the radial coordinate, for
which we sum over the lower 5 ``Walsh-type'' radial modes \cite{geelen2013},
which turns out to be a very reliable method; the inset in Fig.~\ref{fig:scheme}b
shows an example hologram for $\ell=2$. By displaying a series of
vortex holograms on the SLM, we can measure the OAM spectrum $P_{\ell}$.
We assume for now perfect line of sight condition, i.e., $\Delta x\equiv0$. 

\begin{figure}
\includegraphics[width=1\columnwidth]{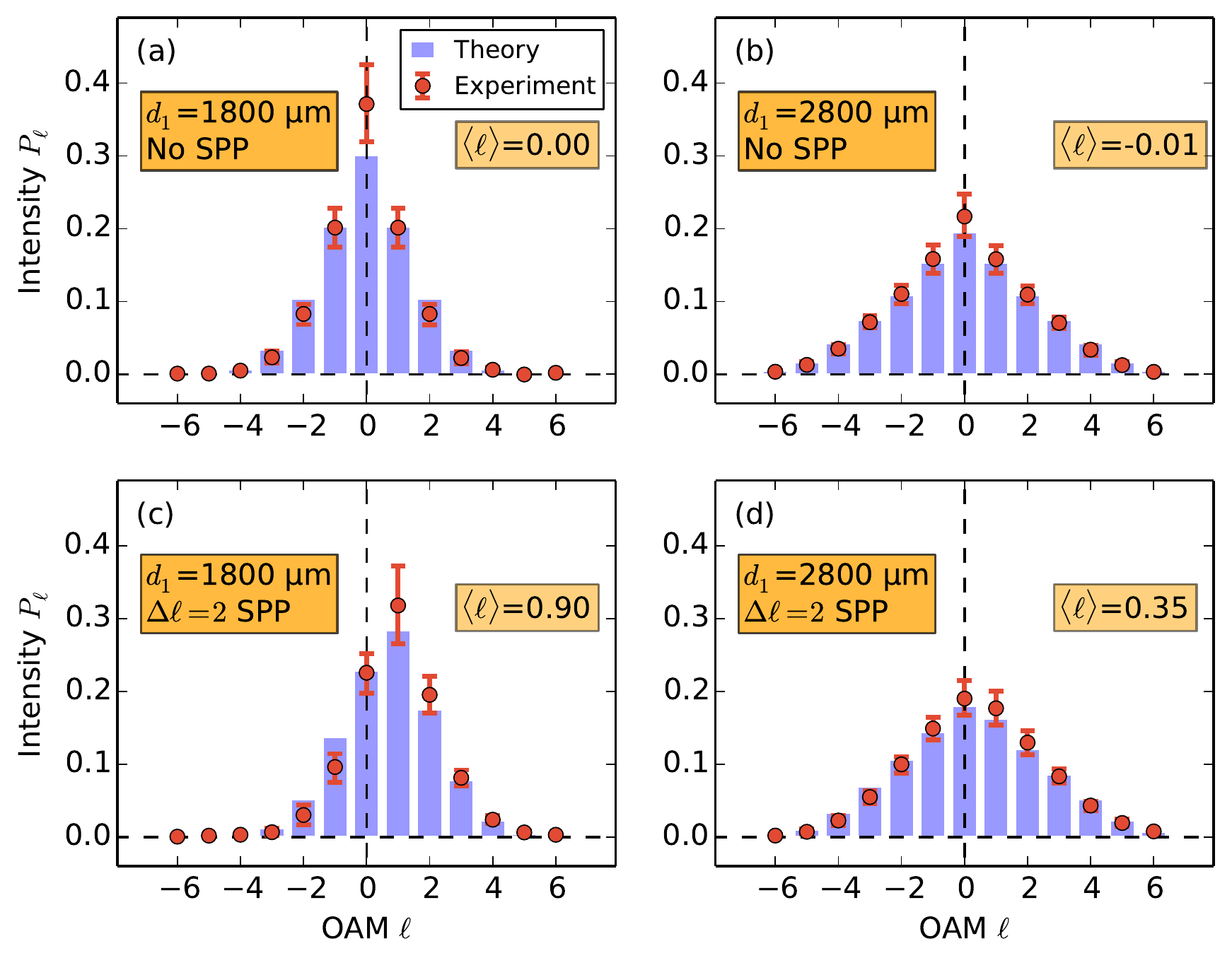}

\protect\caption{\label{fig:oamspectra}Measured OAM spectra (red dots, $P_{\ell}$
versus $\ell$) for different aperture sizes: (a,c) $d_{1}=1800\,\mu$m,
(b,d) $d_{1}=2800\,\mu$m. (a,b) are measured without the SPP, and
(c,d) with a $\Delta\ell=2$ SPP introduced $L_{1}=560$~mm behind
the source aperture $d_{1}$. The experimental error is estimated
from multiple measurements (10\%) and the uncertainty in $d_{1}$
($\pm50\,\mu$m). The bars show the theoretical results (no fit parameters).
Common parameters: $d_{2}=800\,\mu\mathrm{m}$, $L_{2}=315\,\mathrm{mm}$.}
\end{figure}

Fig.~\ref{fig:oamspectra}a and b show the OAM spectra $P_{\ell}$
for two choices of the source diameter $d_{1}$. We recognize the
approximately triangular OAM spectrum, which has been found before
\cite{pires2010}, this is due to the hard edges of the source and
detector aperture, and mathematically based on the Fourier relation
between a squared spherical Bessel function and the triangular function.
We now introduce a SPP with $\Delta\ell=2$ at distance $L_{1}$ from
the first aperture; the resulting OAM spectra are shown in Fig.~\ref{fig:oamspectra}c
and d. Contrary to naive expectation, we do \emph{not} simply observe
a spectrum that is shifted by $\Delta\ell=2$, but we observe a deformed
OAM spectrum where the average shift is smaller than $\Delta\ell=2$.
Further, we see that, the lower the spatial coherence of the source
(i.e., the larger aperture $d_{1}$), the smaller the average OAM
$\langle\ell\rangle=\sum_{\ell}\ell P_{\ell}$. Note that if we use
a single-mode source (not shown), we observe a shift of $\Delta\ell=2$,
as expected. 

How can this be understood? Let us consider briefly two extreme cases:
Clearly, if we place the SPP very close to the spatially incoherent
source (regime A), its action vanishes: The OAM spectrum of the source
is at the SPP position very broad compared to the small $\Delta\ell$
of the SPP, but the detector will receive only a narrow spectrum around
$\ell=0$. Consequently, it detects a nearly unshifted OAM spectrum.
Equally obvious is the opposite extreme regime (B), if the SPP is
placed very close to the detector, it will then detect a $\Delta\ell$-shifted
OAM spectrum, since the action of the SPP can be added to the holograms
used for measurement of the OAM spectrum. Useful cases in astronomy
must lie in between those regimes: Regime A is irrelevant since OAM
carries no additional information, and case B is unrealistic as extremely
close black holes are highly unlikely.

To study the general case, we develop a theory and an intuitive model
that explains our experimentally observed OAM spectrum for arbitrary
aperture sizes, positions of the SPP, and SPP charges $\Delta\ell$.
We could simply calculate the cross-spectral density function $W$
at the detector, and determine from it the OAM spectrum, but this
gives little insight and is very time-consuming. In particular because
we also want to study the case where the observer is not exactly on
the line-of-sight ($\Delta x\neq0$) and no symmetries can be exploited
for simplification \cite{pires2010b}. Our approach here is to model
the incoherent source by a number of Huygens elementary sources at
$\left(x_{1},y_{1}\right)$, each illuminating the SPP with a spherical
wave (which is conceptually related to the method used in \cite{palacios2004}).
For the field directly behind the SPP we obtain ($k=2\pi/\lambda)$:

\begin{align}
E_{s}(x_{s},y_{s}) & =\frac{\exp\left[ikR_{s}\right]}{R_{s}}\cdot\exp\left(i\Delta\ell\phi\right)\\
\mathrm{with} & \; R_{s}^{2}=(x_{s}-x_{1})^{2}+(y_{s}-y_{1})^{2}+L_{1}^{2}\nonumber 
\end{align}

We then propagate this field to the detector ($E_{2}$) numerically,
using the well-known Huygens-Fresnel principle:

\begin{align}
E_{2}(x,y) & =\int dx_{s}dy_{s}E_{s}(x_{s},y_{s})\exp(i\frac{2\pi}{\lambda}R)/R\label{eq:e2}\\
\mathrm{with} & \; R^{2}=(x-x_{s})^{2}+(y-y_{s})^{2}+L_{2}^{2}\nonumber 
\end{align}

For each elementary source at $\left(x_{1},y_{1}\right)$, the OAM
spectrum $P_{\ell}(x_{1},y_{1})=\int_{0}^{d_{2}/2}r\, dr\left|\int d\phi E_{2}(r,\phi)\exp(i\ell\phi)\right|^{2}$
is calculated and summed up incoherently for all sources: $P_{\ell}=\int\mathrm{d}x_{1}\mathrm{d}y_{1}E_{1}(x_{1},y_{1})P_{\ell}(x_{1},y_{1})$.
In case $\Delta x=0$ (exactly in line-of-sight), due to symmetry,
we can avoid one integral and only propagate Huygens sources emerging
from along a radius of the source (e.g., for $0<x_{1}<d_{1}/2$ with
$y_{1}=0$), and add a radial factor in summing up the OAM spectra
as $P_{\ell}=\int_{0}^{d_{1}/2}\mathrm{d}x_{1}\, x_{1}P_{\ell}(x_{1},y_{1}=0)$.
The result of this numerical simulation is compared in Fig.~\ref{fig:oamspectra}
to the experimental data; we see good agreement and good reproduction
of the OAM spectrum deformations. 

But for applications of OAM, the precise shape of the spectrum is
often not relevant. Hence, we focus in the following on the \emph{mean
OAM} $\langle\ell\rangle$, which is directly measurable experimentally
\cite{piccirillo2013}. We find here that the mean OAM is a robust
quantity that allows further approximations of Eq.~\ref{eq:e2} %
\footnote{In the intermediate regime, we found that the Fresnel propagator can
be completely removed without strongly modifying \unexpanded{$\langle\ell\rangle$}.%
} and also to derive a simple model.

\begin{figure}
\includegraphics[width=1\columnwidth]{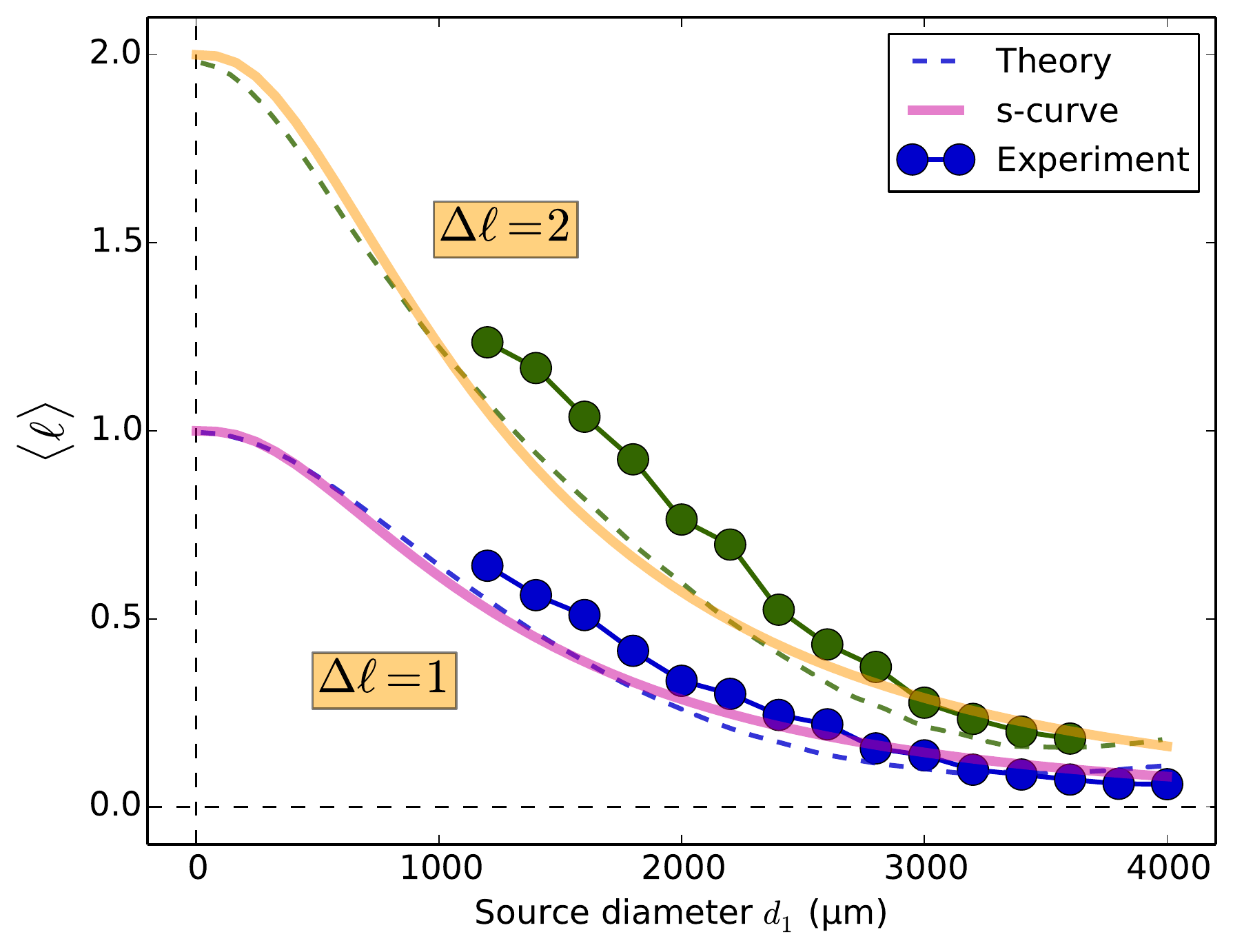}

\protect\caption{\label{fig:modelexp}The detected mean OAM as a function of the aperture
size $d_{1}$ for $\Delta\ell=1$ and $\Delta\ell=2$ spiral phase
plates placed $L_{1}=56$~cm behind the first aperture ($d_{2}=800\,\mu\mathrm{m}$,
$L_{2}=315\,\mathrm{mm}$). The experimental data (symbols) agree
well to the theory (Eq.~\ref{eq:e2}) and the simple s-curve model.}
\end{figure}

First, we investigate the dependency of the OAM mean $\langle\ell\rangle$
on the diameter of the source $d_{1}$. In an astrophysical context
this corresponds to the diameter of the source star, which determines
the degree of spatial coherence at the position of the light-twisting
object. In Fig.~\ref{fig:modelexp} we compare $d_{1}$-dependent
experimental data with the theory (Eq.~\ref{eq:e2}). We see that
only for very small source diameter, the full OAM shift introduced
by the SPP is also detectable at the detector. We have measured this
for two spiral phase plates with $\Delta\ell=1,2$ in Fig.~\ref{fig:modelexp},
the similar shape of the curves suggests that the charge of the SPP
is simply a scaling parameter, at least for low $\Delta\ell$, we
therefore normalize in the following the observed mean OAM by $\Delta\ell$.

How can we estimate the detected mean OAM in terms of a simple model?
We found that we simply have to compare the coherence length $L_{i}^{c}=1.22\lambda L_{i}/d_{i}$
of the source star $L_{1}^{c}$ and that of the backpropagated detector
$L_{2}^{c}$ at the position of the light-twisting object as follows:

\begin{equation}
\mathcal{F}=\frac{L_{1}^{c}}{L_{1}^{c}+L_{2}^{c}}=\frac{d_{2}L_{1}}{d_{2}L_{1}+d_{1}L_{2}},\label{eq:finalF}
\end{equation}

\begin{figure}
\includegraphics[width=1\columnwidth]{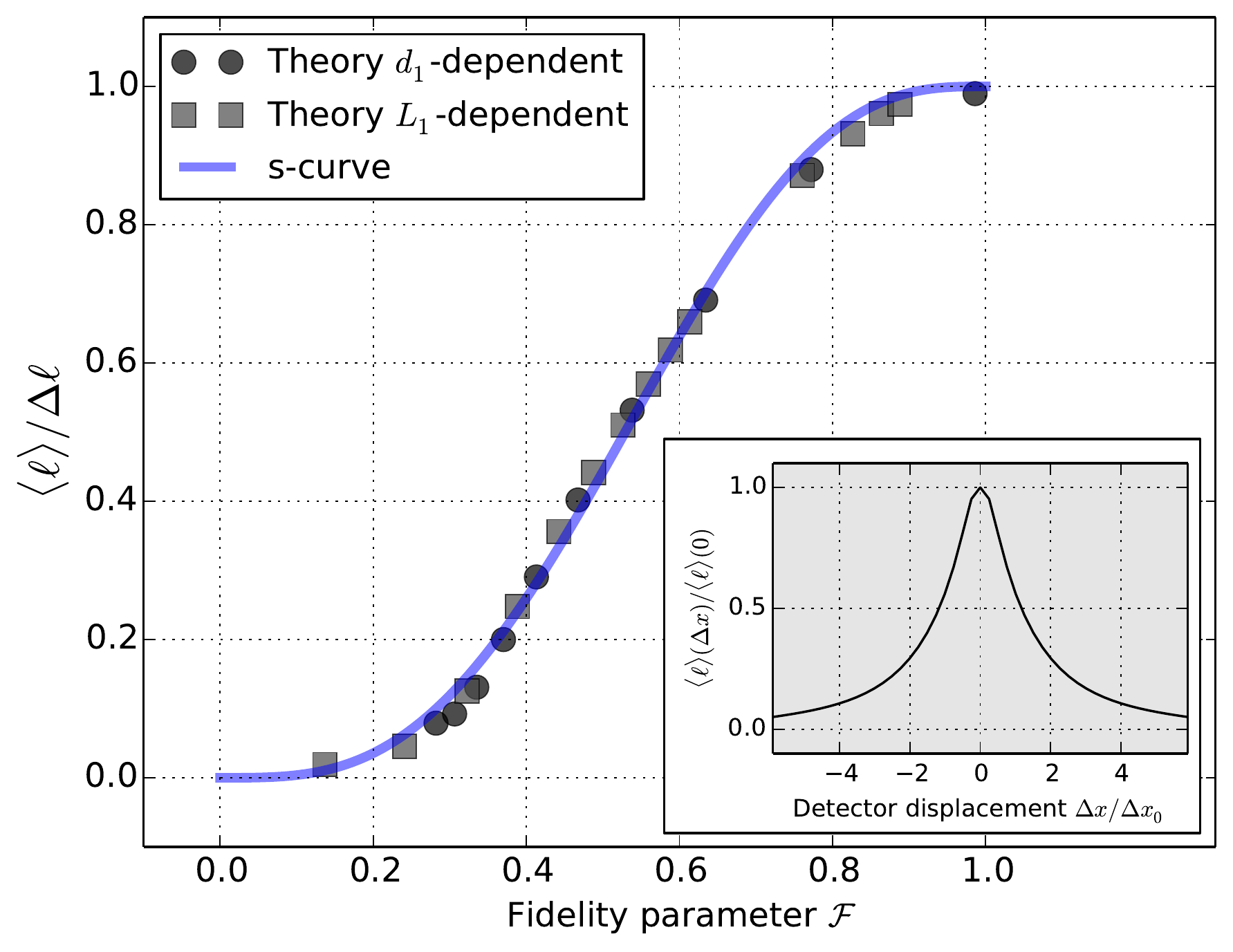}

\protect\caption{\label{fig:scurve}Detected normalized mean OAM for $d_{1}$ and $L_{1}$
dependent calculation (symbols), plotted as a function of the parameter
$\mathcal{F}$. Both calculations lie on a s-shaped curve that is
nicely represented by the incomplete Beta-function (curve). Inset:
Calculated mean OAM as a function of detector displacement in multiples
of $\Delta x_{0}=\sqrt{d_{1}d_{2}L_{2}/L_{1}}$, relative to the exact
line-of-sight condition $\Delta x=0$.}
\end{figure}

To confirm this choice, Fig.~\ref{fig:scurve} shows the previous
$d_{1}$-dependent calculation together with new $L_{1}$-dependent
calculations. We find that the data points lie on a common curve.
This confirms that all distance ($L_{1},L_{2}$), diameter ($d_{1},d_{2})$,
and SPP charge ($\Delta\ell$) dependencies can nicely be mapped by
$\mathcal{F}$ onto a sigmoid-shaped curve. A s-shaped curve that
fits best and has fewest parameters is the incomplete Beta function,
we obtain for the mean OAM $\langle\ell\rangle=\Delta\ell\cdot B(\mathcal{F},\,3.5,\,3.1)$,
see Fig.~\ref{fig:scurve}. This agrees perfectly with the intuitive
picture that we have presented in the beginning, namely that if the
SPP is close to the source ($L_{1}\ll L_{2}$, assuming $d_{1}\approx d_{2}$),
its action disappears because the field at that position is highly
incoherent. On the other hand, very close to the detector ($L_{1}\gg L_{2}$),
the OAM shift $\Delta\ell$ from the SPP is fully reflected in the
detected mean of the OAM spectrum $\langle\ell\rangle$, independent
of the spatial coherence of the field there. An estimation of the
fidelity parameter Eq.~\ref{eq:finalF} could also be derived from
relating the detector diameter $d_{2}$ to the coherence singularity
diameter $d_{2}^{c}=d_{1}L_{2}/L_{1}$ at the position of the detector
\cite{palacios2004}. This explains why the expression \ref{eq:finalF}
is not wavelength dependent: for longer wavelengths, diffraction only
reduces the over all intensity and does not modify the ratio between
the individual detected OAM modes.

Now, we discuss the possibility of using light's OAM in astronomical
observations of massive, space-distorting objects such as rotating
black holes that might twist light \cite{feng2001,tamburini2011}.
As an example, we consider the supermassive Kerr black hole at the
center of our galaxy, Sagittarius A{*} (Sgr~A{*}) with a Schwarzschild
radius of $R_{s}=1.27\times10^{12}$~m, where it is reasonable to
assume that all light that we receive is modified \cite{tamburini2011}.
First, is observation at radio frequencies or with visible light advantageous?
Although the wavelength $\lambda$ disappeared from Eq.~\ref{eq:finalF},
it affects the overall detected intensity (which scales as $\left[d_{2}^{2}/\left(L_{2}/k\right)\right]^{|\Delta\ell|}$,
see below). In these terms, most radio telescope arrays are comparable
to visible-light interferometric telescopes, we select as an example
the next-generation Magdalena Ridge Observatory Interferometer (MROI)
with $d_{2}=340$~m baseline, operating at $\lambda=600$~nm with
an angular resolution of $1.8\times10^{-9}$~rad. This telescope
would receive from this area {[}at $L_{2}=25.9\times10^{3}$~light
years (ly) distance{]} up to $N_{OAM}=2.3$ optical OAM modes \cite{pires2010}.
However, as we have shown, it is highly unlikely that such a large
area will be illuminated spatially coherently. It is more realistic
to assume that a sun-like star or pulsar is illuminating Sgr~A{*}
from behind, exactly on the line of sight, we choose $d_{1}=1.4\times10^{9}$~m.
From Eq.~\ref{eq:finalF} we see that, to observe on earth a mean
OAM $\langle\ell\rangle=\Delta\ell/2$ (i.e., 50\% shift), we have
the extreme requirement that distance $L_{1}>d_{1}L_{2}/d_{2}\approx10^{11}\,\mathrm{ly}$,
which is impossible. However, more recent studies suggest that a large
number of smaller black holes exist in every galaxy such as ours;
for instance, for a black hole in the Orion nebula at $L_{2}=1500$~ly
distance \cite{subr2012}, an illuminating object at only $L_{1}=6\cdot10^{9}$~ly
would be needed. If we can detect a 1\% OAM shift $\langle\ell\rangle=0.01\cdot\Delta\ell$,
an illuminating star at $L_{1}=6\cdot10^{7}$~ly would suffice.

Up to now, the source, the light-twisting object, and the observer
were considered to be perfectly on a line. For large detectors, it
is known that misalignment leaves the mean OAM unchanged, only the
variance of the OAM spectrum increases \cite{gibson2004,vasnetsov2005,zambrini2006,loffler2012b}.
Here, however, due to the limited aperture of our detector, we find
that the detected mean OAM $\langle\ell\rangle$ is reduced by transverse
misalignment $\Delta x$, calculated by modifying $P_{\ell}(x_{1},y_{1})$
below Eq.~\ref{eq:e2}. For a coherent light source, if the detector
is displaced by more than its diameter, i.e., $\Delta x\gg d_{2}$,
the detected mean OAM $\langle\ell\rangle(\Delta x)$ vanishes. This
would render OAM useless for astronomy because of typically very large
transverse speeds, e.g., the earth is moving at $v=30\,\mathrm{km\; s^{-1}}$
around the sun. However, for partially coherent light, we find a different
scaling parameter $\Delta x_{0}=\sqrt{d_{1}d_{2}L_{2}/L_{1}}$, see
the inset in Fig.~\ref{fig:scurve}, where the calculated mean OAM
normalized to zero displacement $\langle\ell\rangle(\Delta x)/\langle\ell\rangle(\Delta x=0)$
is shown. For the case of a rotating black hole in the Orion nebula
(see above, for 1\% OAM shift), we obtain $\Delta x_{0}=3450$~m,
much larger than the detector size; reduced spatial coherence is actually
advantageous here. The observation of $\langle\ell\rangle$-transients
due to relative motion opens up a novel possibility to find, for instance,
close-by black holes as generally, non-zero total OAM is expected
to be strictly absent.

Finally, we briefly discuss a well-known but often ignored fact: During
propagation from the light-twisting object (SPP) to the detector,
the vortex core expands due to diffraction, and if its effective diameter
is large compared to the detector, the detected intensity is much
lower compared to the case without SPP. In the astronomy case, the
earth-bound observer is certainly always in this regime and for estimatation
of this effect, we find that the size of the dark core of an OAM mode
scales as $\left(1+\ell/2\right)\sqrt{L_{2}/k}$ (in agreement with
the $\ell=1$ case in \cite{khonina1992}). The integrated intensity
captured by a detector of diameter $d_{2}$ is %
\footnote{This is calculated based on the $\sim r^{|\ell|}$ amplitude close
to the vortex core and confirmed via numerical simulations. Mind that
this is different for exact Laguerre-Gaussian beams.%
}, relative to the plane-wave case, $I_{det}(\ell)/I_{det}(0)=\pi\left(d_{2}^{2}k/4z\right)^{\left|\ell\right|}/\left(1+\left|\ell\right|\right)$.
For the case of Sgr~A{*} observed with the MROI, this ratio is approximately
$10^{-9\left|\ell\right|}$, suggesting that OAM is only suitable
for observation of much closer light-twisting objects, because an
OAM astronomer sits always in the ``shadow of the phase singularity''.
But mind that this is always the case if we can resolve an image of
an object far away.

In conclusion, we have found experimentally and theoretically that
insertion of a light-twisting object, such as a spiral phase plate
with charge $\Delta\ell$, in light with reduced spatial coherence
results in detected OAM spectra which depend strongly on the position
of this object. This contrasts with the well-known coherent case,
where simply a displacement of the OAM spectrum by $\Delta\ell$ occurs.
We have derived a simple parameter (Eq.~\ref{eq:finalF}) for the
mean of the detected OAM spectrum $\langle\ell\rangle$ as this is
key for applications. For observation of a nonzero OAM shift, it is
required that as few as possible modes from the source illuminate
the light-twisting object, and as many modes as possible are detected
from it by the observer. With this, we have assessed the use of OAM
in astronomy and find that with current technology, only close-by
light-twisting objects are within reach; additionally, low light levels
and line-of-sight mismatch must be taken into account seriously. However,
the detrimental effects of line-of-sight misalignment might be smaller
for spatially incoherent sources, thus first detection of astronomical
OAM could actually be facilitated in this case. It would be interesting
to study the influence of gravitational (micro-) lensing \cite{ciufolini2002},
possibly by the black hole itself \cite{bozza2010}, on light collection
and the line-of-sight criterion; wave-optical studies of these cases
are needed.

\section*{Acknowledgments}

We acknowledge Christoph Keller (Leiden) for fruitful discussions
and financial support by NWO (grant number 680-47-411).

\bibliographystyle{apsrev4-1}
\bibliography{bibliography}

\end{document}